# Nanopattern on Carbon and by Carbon

**S. Bhattacharjee[1], P. Karmakar[2*], V. Naik[2], A. K. Sinha[1] and A. Chakrabarti[2]**

[1]UGC-DAE Consortium for Scientific Research, III/LB-8, Saltlake, Kolkata -700098, India

[2]Variable Energy Cyclotron Center, I/AF, Bidhannagar, Kolkata -700064, India

## Abstract

We have reported nanopattern formation on carbon thin film and Si(100) surfaces by low energy inert and carbon ion beams. It is interesting to observe the role of carbon as target as well as projectile for nano patterning. Using carbon thin film as target, nano patterns of carbon are formed by inert ($Ar^+$) and self ($C^+$) ion bombardment, whereas carbon ion beam is used to form well ordered Si nano ripple structure in a cost effective way where implanted carbon plays an important role to form Si ripple in relatively lower fluence than the inert projectile.

---

* Corresponding author, E-mail: prasantak@vecc.gov.in

## Introduction

Self organization ion beam induced surface nanopatterning has received considerable theoretical and experimental attention due to its potential applications and challenges to explore its complex physical mechanism [1,2]. The presence of impurities on the surface, multi elemental surface, projectile or environment induced contamination on the surface are reported as the initiator of the nano patterns [3-6]. Then again, there have been a lot of interests in carbon films in the last two decades because of their beneficial chemical and physical properties like high chemical inertness, diamond like physical properties and tribological properties suitable for industrial usage [7,8]. Their unique properties can be attributed not only to the special and interesting properties of the microstructures but also their surface morphology. Also it is observed that the morphology on diamond like carbon film affects the properties of the films such as friction coefficient, optical characteristics as well as electronic properties [9]. Ion induced self organized ripple formation on graphite and diamond surfaces are already reported [10,11]. Nanostructured surface layers of titanium carbide and tungsten carbide were prepared on tetrahedral amorphous carbon film using the surfactant sputtering technique [12]. Experimental and simulation studies of low energy $C^+$ and $Ar^+$ ions bombardment on Si pitch grating are also reported recently [13-15]. However, ripple structure formation on carbon thin films by Ar and self ion bombardment has not been compared yet. Moreover, the role of implanted carbon on ripple formation has not been explored.

In this work we took two sets of samples, in the first sets carbon films are prepared on Si(100) surface and in the another set pure Si(100) surface are taken as target. Both are bombarded with keV energy $C^{1+}$ ions and $Ar^{1+}$ ions and studied the growth of ripple structures on carbon films and Silicon surface due to inert ($Ar^+$) and self ($C^+$) ion irradiation. In case of Si target the role of implanted carbon on nanoripple formation is compared with the implanted inert Ar ions.

**Experimental**

The commercially available polished Si (100) samples are initially cleaned and degreased with trichloroethylene in ultra sonic cleaner and then washed with methanol and distilled water. The carbon film of about 100 nm thickness are then deposited on the half of the Si(100) substrates by thermal vapour deposition. The other half are stored in vacuum. The samples (Carbon film and Si(100)) are bombarded with 8 keV $C^{1+}$ and $Ar^{1+}$ ion beams at an incidence angle of $60^0$ with respect to the surface normal. The fluence was varied from $5\times10^{16}$ ions/cm$^2$ to $7\times10^{17}$ ions/cm$^2$. The ion beam was extracted from 6.4 GHz ECR ion source of the Radioactive Ion Beam Facility at Variable Energy Cyclotron Centre Kolkata (VECC) [16]. The ion current density was ~ 5μA/cm$^2$ measured with current integrator and the chamber pressure was $3\times10^{-7}$ mbar during the experiment.

The surface morphology of the carbon films and Si(100) samples before and after implantation were examined in air using Bruker Atomic Force Microscopy (AFM), MM V at VECC, Kolkata. Also the samples were characterized using X- ray photoelectron spectroscopy (XPS) to get the relative composition of the constituents in the surface region along with the identification of the element presents and their chemical states. The XPS instrument is from Oxford Applied Research with monochromatic X-ray source of Mg($k_\alpha$) (1253.6eV) and 150mm mean radius hemi spherical analyzer. The pressure in the chamber was $8\times10^{-10}$ mbar during irradiation with X-rays.

## Results and Discussions

The surface morphology of the virgin carbon film and the bombarded samples is shown in Fig.1. Fig.1 (a) shows the AFM image of the unbombarded carbon film deposited on Si(100) substrate. The surface topography of carbon films bombarded with 8 keV $C^{1+}$ ions at $60^0$ with respect to the surface normal for fluences $5\times10^{16}$ ions/cm$^2$ to $7\times10^{17}$ ions/cm$^2$ are shown in Fig. 1 ($b_1$-$b_5$). Similarly, Fig.1 ($c_1$-$c_5$) shows the images of the films bombarded with $Ar^{1+}$ ions for the same variation of ion fluence. It is observed from the topographic AFM images that the ripples like patterns are formed on the film surface with increasing ion fluence. For $C^{1+}$ ion bombardment the initiation process started at the fluence $5\times10^{16}$ ions/cm$^2$ and with the increase of fluence the ripple patterns are developed gradually. A good quality ripple structure is observed at the fluence of $7\times10^{17}$ ions/cm$^2$ with the ripple wavelength ~ 100 nm. But in the case of $Ar^{1+}$ ions similar ripple patterns with wavelength ~ 100 nm are observed at a lower fluence of $2\times10^{17}$ ions/cm$^2$ only. With the further increase of $Ar^{1+}$ fluence the ripple patterns are randomized leading to kinetic roughening.

The ion beam induced nanopatterning is described by the instability generation on the surface during ion bombardment. The ion energy and momentum transfer to the surface atoms leads to mass redistribution [17-19] and sputtering which cause perturbation on the surface morphology. With continuous ion irradiation, the perturbation grows and generates structure on the surface depending on the incident beam and target parameters. The BH model [20] or extended BH model, based on Sigmund's sputtering theory [21] is commonly used to describe the ripple formation at oblique angle ion bombardment. An initial topographic perturbation on the surface and ellipsoidal shape of the collision cascade during ion bombardment are assumed in this model. The perturbation grows with ion bombardment and generates local curvature. Because of the geometry more energy from the collision cascades reaches to the concave curvature (valleys) than the convex curvatures (hills) on the surface and therefore, the preferential sputtering of the valleys generate instability. In addition to the sputtering induced instability, thermal diffusion and ion beam induced mass redistribution play the crucial role on ripple formation [18,20].

In the present case, as grown C thin films has initial rough morphology (Fig.1 a), therefore, the initial perturbation is readily available, and thus ion bombardment at $60^0$ leads to the formation of perpendicular ripple structure following the curvature dependent sputtering and mass redistribution. Similar ripples were also observed in ZnO film [22]. It is to be pointed out here that for initially flat surfaces, stochastic nature of ion beam generates the initial perturbation. Preferential sputtering, impurities on the surface or deliberate incorporation of initial roughness may also supply the initial perturbation and therefore, speed up the ripple formation during ion bombardment [3,23].

Fig. 1(d) and (e) show the variation of roughness and wavelength of the structures with the fluence of $C^{1+}$ and $Ar^{1+}$ ions. For $C^{1+}$ ions it is observed that rms roughness is always less and increases slowly with the ion fluence compared to Ar ions (Fig. 1 d). When the wavelength of the structures are compared, it is observed that in case of $Ar^+$ ion the wavelength increases slowly at the initial stages and then faster at around fluence $2\times10^{17}$ ions/cm$^2$ but ripple wavelengths are almost constant for $C^+$ ions (Fig.1. e). After the $Ar^+$ fluence of $2\times10^{17}$ ions/cm$^2$ the sharp increase of wavelength is due to the kinetic roughening of the sample. The topography images (Fig.1$c_4$-1$c_5$) also shows that after the fluence of $2\times10^{17}$ ions/cm$^2$ the ripple patterns are distorted. The observed decrease of roughness at $Ar^+$ fluence of $7\times10^{17}$ ions/cm$^2$ is due to the erosion loss of the carbon film as was observed earlier in case of ZnO films [22] and Pt films [24].

Due to higher mass of Ar ions, the penetration of the ion is less than the same energy C ion but effective energy loss per unit length is higher, which leads to higher sputtering and target atom redistribution compared to C ion. The total energy loss per unit length (stopping power) is 962 eV/nm for $Ar^{1+}$ and 332 eV/nm for $C^{1+}$ on carbon. The sputtering yield for Ar projectile is 5.35 atoms/ion compared to 1.76 atoms/ion for $C^{1+}$ ions [25]. Following the Carter - Vishnyokov effect (CV effect) [26], Madi et al. [18] predicted that the mass redistribution induced instabilities play the only role for pattern formation. Recently Bobes et al. [27] showed that in addition to mass redistribution, curvature dependent sputtering (BH model) is also vital for nano patterning. Here, Ar impact induced higher sputtering and mass redistribution generates more instability on carbon thin film than self ion ($C^+$) bombardment. Therefore, early appearance of ripple as well as higher surface roughness and ripple wavelength for $Ar^+$ bombarded carbon films as observed is consistent with both CV effect and BH model.

The XPS measurement of carbon films on Si(100) before and after ion bombardment is consistent with the AFM data. The Si peaks don't appear for as deposited carbon films and only C(1s) and O(1s) peaks are visible. The intensity of the C(1s) is found to decrease when the carbon film is bombarded either by $C^{1+}$ or $Ar^{1+}$ ions. The faster decrease of carbon intensity is observed for Ar ion bombardment as shown in Fig.2(a). Si peaks appears for Ar ion bombarded carbon film, whereas no such Si peaks are observed for bombardment with C ions Fig 2(b). The higher roughness of $Ar^+$ bombarded carbon surface and decrease of roughness after fluence $5\times10^{17}$ ions/cm$^2$ (Fig 1.d) indicate that the sputtering yield is higher than C ion and after a certain fluence the Si surface is exposed for $Ar^+$ ion bombardment. Thus, the higher sputtering yield and higher target atom redistribution capability of $Ar^+$ ions leads to the faster ripple formation compared to $C^+$ ions. Due to inert nature of Ar ion and self bombardment of carbon film by C ion no chemical effect is expected for ripple formation.

Fig.3 shows the chemical effect on nano structure formation where the target is changed from carbon film to pure Si (100). Fig. 3(b1-b2) and (c1-c5) shows the AFM images of the morphology developed on Si surfaces due to bombardment of $Ar^{1+}$ and $C^{1+}$ ions respectively. In case of $C^+$ ion bombardment the smooth Si surface start to increase its rms roughness; at fluence greater than $5\times10^{17}$ ions/cm$^2$ the ripple structures appear and the surface roughness increases sharply. The wavelength of the ripple structure is also found to increase with the ion fluence (Fig.3e). In case of Ar ion bombardment no ripple is formed in the present ion energy even at very high fluence of $2\times10^{18}$ ions/cm$^2$. Initial bombardment of $C^{1+}$ ions at fluence less than $5\times10^{17}$ ions/cm$^2$ generates only random roughness on the surface, where sputtering induced initial height perturbation is not enough to form the sufficient instability on the surface to form the ripple structure. At larger fluence the implanted carbon ions change the chemical nature of the Si surface. The chemical shift of Si (2p) peak towards higher binding energy ($\Delta E$ = 1.255) (shown in Fig 3d) indicates the SiC formation. With increasing carbon implantation binding energy of the Si (2p) signal shifts towards the higher energy. The charge transfer from the silicon atoms to the more electronegative implanted carbon atoms leads to the shift of Si (2p) peaks towards higher binding energy. Yan et.al observed the similar shift of Si (2p) peaks from $C^{1+}$ implanted Si sample [28]. Poudel et al. [29] reported the SiC formation from XPS measurement of C ion implanted Si at elevated temperature. $Ar^{1+}$ ion implanted

Si surface does not show any chemical shift of the Si (2p) peaks. Therefore Ar ion bombardment does not generate sufficient instabilities to form the ripple structure on the surface. No structure formation on Si by Ar bombardment in this energy range is also reported earlier [6].

In the presence of initial perturbation in terms of initial roughness ripples are formed on carbon thin films by both $Ar^+$ and $C^+$ ions bombardment. However, for higher sputtering and mass redistribution effect of $Ar^+$ on carbon thin film than the same energy self ions ($C^+$), more effective ripple formation is observed for $Ar^+$ bombardment. But no ripples are formed by $Ar^+$ ion bombardment on initially smooth Si surface whereas $C^+$ ion having lower sputtering yield and mass redistribution efficiency than $Ar^+$ leads to ripple formation on smooth Si(100) surface because of chemical reaction of carbon with Si. Therefore it can be concluded that surface chemistry play the most important role in instability formation for nanopatterning. In view of that, pattern formation on solid surfaces by low energy ion bombardment is possible in all cases either with suitable initial perturbation or allowing chemical reaction during ion bombardment where ion induced kinematics in not sufficient to generate instabilities.

The authors would like to thank Prof. Satyaranjan Bhattacharya and Mr. Shyamal Mondal for XPS measurements. S. Bhattacharjee is thankful to CSIR for financial support.

**Figure Captions**

**Fig.1** AFM images of the (a) virgin carbon film, ($b_1$–$b_5$ ) ripple like pattern form on carbon film due to $C^{1+}$ bombardment and ($c_1$ – $c_5$) due to $Ar^{1+}$ bombardment for ion fluence varying from $5\times10^{16}$ ions/cm$^2$ to $7\times10^{17}$ions/cm$^2$ at an incidence angle of $60^0$. Arrows indicate the direction of the ion beams. (d) Shows the variation of roughness and (e) the wavelength with the ion fluence for $C^{1+}$ and $Ar^{1+}$ ion bombardment.

**Fig 2**(a) and (b) Shows the XPS peak of C(1s) and Si(2p) for virgin carbon film on Si, $C^{1+}$ ion implanted on C/Si and also $Ar^{1+}$ ion implanted on C/Si sample.

**Fig.3** AFM images of the (a) virgin Si(100) surface, (b1) $Ar^+$ ion bombarded Si surface for the fluence of $1\times10^{18}$ ions/cm$^2$ and (b2) $2\times10^{18}$ ions/cm$^2$,
(c1-c5) $C^+$ ion bombarded on Si surface for the fluence varying from $5\times10^{16}$ions/cm$^2$- $7\times10^{17}$ions/cm$^2$ at $60^0$ incidence angle with respect to the surface normal . Arrows indicates the direction of the ion beams.
(d) Shows the XPS peak for Si(2p) for the virgin Si sample and C implanted Si surface.(e) Shows the variation of roughness and ripple wavelength with the ion fluence.

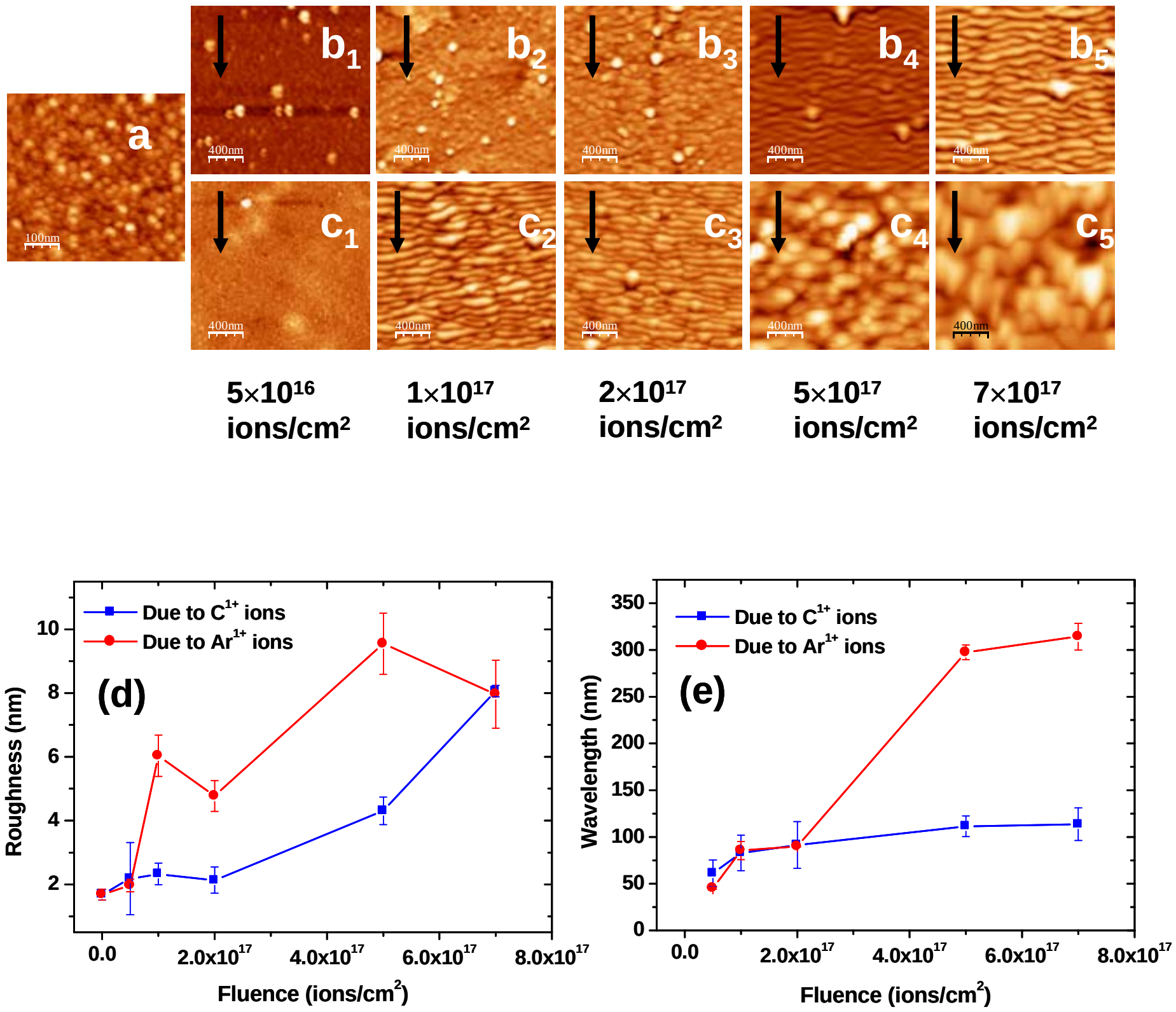

**Figure 1**

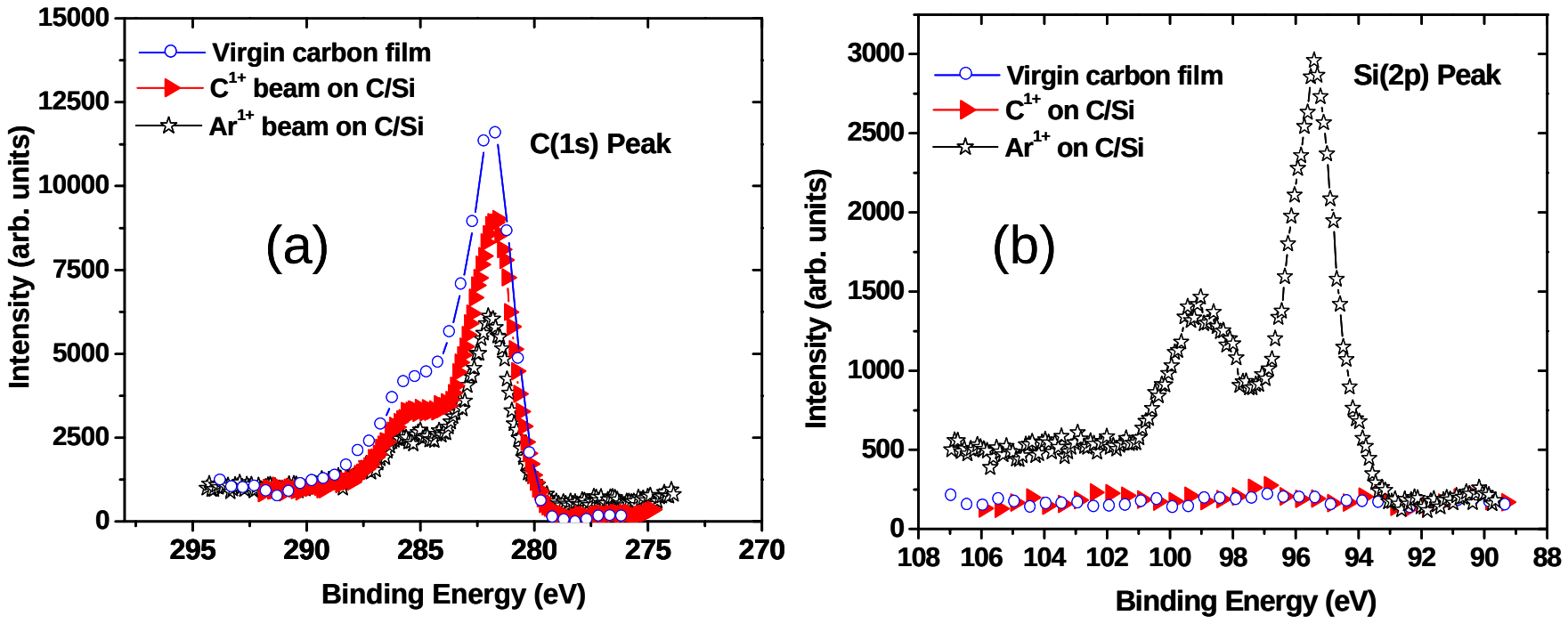

**Figure 2**

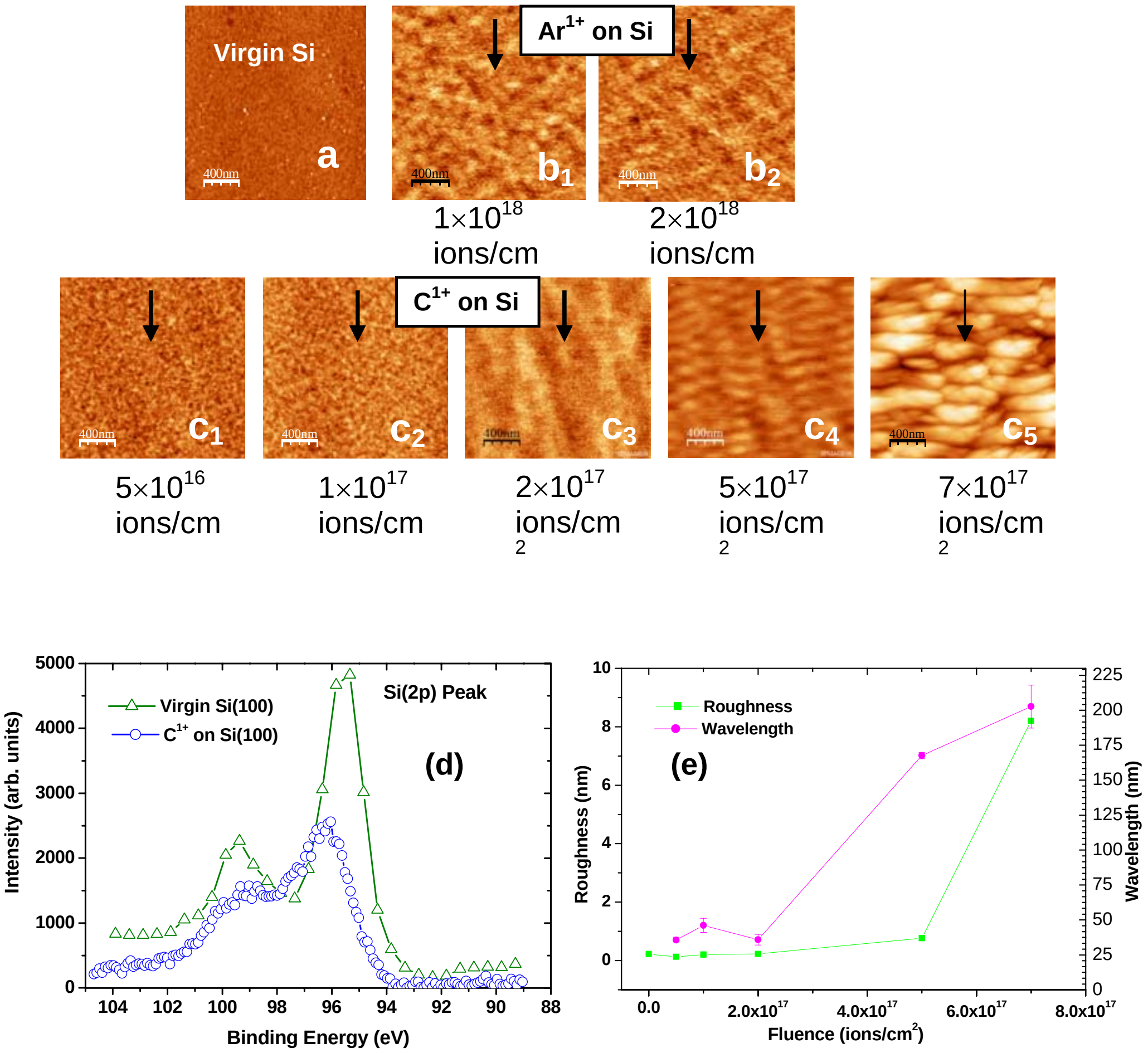

**Figure 3**